# Influence of study type on Twitter activity for medical research papers

Jens Peter Andersen[1] and Stefanie Haustein[2]

[1] *jepea@rn.dk*
Aalborg University Hospital, Medical Library, DK-9000 Aalborg (Denmark)

[2] *stefanie.haustein@umontreal.ca*
École de bibliothéconomie et des sciences de l'information, Université de Montréal, Montréal (Canada)

**Abstract**
Twitter has been identified as one of the most popular and promising altmetrics data sources, as it possibly reflects a broader use of research articles by the general public. Several factors, such as document age, scientific discipline, number of authors and document type, have been shown to affect the number of tweets received by scientific documents. The particular meaning of tweets mentioning scholarly papers is, however, not entirely understood and their validity as impact indicators debatable. This study contributes to the understanding of factors influencing Twitter popularity of medical papers investigating differences between medical study types. 162,830 documents indexed in Embase to a medical study type have been analysed for the study type specific tweet frequency. Meta-analyses, systematic reviews and clinical trials were found to be tweeted substantially more frequently than other study types, while all basic research received less attention than the average. The findings correspond well with clinical evidence hierarchies. It is suggested that interest from laymen and patients may be a factor in the observed effects.

**Conference Topic**
Altmetrics

**Introduction**
In the context of altmetrics, defined as "the study and use of scholarly impact measures based on activity in online tools and environments" (Priem, 2014, p. 266), Twitter has been identified as one of the most interesting and widely-used data sources (Costas, Zahedi, & Wouters, 2014; Thelwall, Haustein, Larivière, & Sugimoto, 2013). Although restricted by brevity—a tweet is limited to 140 characters—Twitter is at the heart of the altmetrics idea to enable a broader scope for impact assessment beyond citation impact. As Twitter is used widely and particularly outside of academia by currently 284 million monthly active users[1], tweets mentioning scientific papers are hoped to capture use by the general public and thus societal impact. Initially suggested as predictors of future citations and thus early indicators of scientific impact (Eysenbach, 2011), more recent large-scale empirical studies suggest that tweets are more likely to reflect online visibility including some social and scientific impact but also self-promotion and buzz (Costas et al., 2014; Haustein, Larivière, Thelwall, Amyot, & Peters, 2014; Haustein, Peters, Sugimoto, Thelwall, & Larivière, 2014). The most tweeted documents seem to attract a lot of online attention rather due to humorous or curious topics than their scientific contributions, often fitting "the usual trilogy of sex, drugs, and rock and roll" (Neylon, 2014, para. 6).

Various, mostly quantitative, studies have shown, with respect to scientific papers, that—after the reference manager Mendeley—Twitter is the altmetrics data source with the second-largest prevalence and it is constantly increasing to currently more than one fifth of 2012 papers being tweeted (Haustein, Costas, & Larivière, 2015). Correlation studies provide evidence that tweets and citations measure different things (for example, Costas et al., 2014; Haustein, Larivière, et al., 2014; Haustein, Peters, et al., 2014; Priem, Piwowar, & Hemminger, 2012; Thelwall et al., 2013; Zahedi, Costas, & Wouters, 2014). The latest research shows that Spearman correlations with citations for 2012 papers in Web of Science are low at $\rho=0.194$ for all 1.3 million papers and $\rho=0.148$ excluding untweeted papers. Beyond the particular differences of Twitter coverage and density between scientific disciplines, research fields and journals reported by various studies (Costas et al., 2014; Haustein, Larivière, et al., 2014; Haustein, Peters, et al., 2014; Zahedi et al., 2014), Haustein et al. (2015) also identified large variations between document types deviating from patterns known for citations. For example, news items and editorial material, which are usually considered non-citable items (Martyn & Gilchrist, 1968), are the most popular types of journal publications on Twitter, showing a tendency of increasing Twitter impact for brief and condensed document types. A study based on a random sample of 270 tweets to scientific papers found that the majority of tweets contained either the paper title or a summary, did not attribute authorship and had a neutral sentiment, while 7% were self-citations (Thelwall, Tsou, Weingart, Holmberg, & Haustein, 2013). Other findings suggest that automated diffusion of article links on Twitter plays a role as well (Haustein, Bowman, et al., 2015).

---
[1] https://about.twitter.com/company

Although these findings provide more evidence that the mechanisms behind tweeting a paper are different from those citing it, the meaning of tweets to scientific papers as well as the role of Twitter in scholarly communication are still unclear, not in the least due to the difficulty to identify 'tweeter motivations' based on 140 characters. This study aims to contribute to a better understanding of tweets as impact metrics by analysing the type of content that is distributed on Twitter. We propose that certain types of articles appeal more to the public than others, for example, because of their potential impact on health issues and everyday life or due to the fact that they are written in a certain way. Previous research has suggested that certain medical study types have a larger citation potential than others (Andersen & Schneider, 2011; Kjaergard & Gluud, 2002; Patsopoulos, Analatos, & Ioannidis, 2005), likely because they are more useful to the research community. In the context of Twitter, medical papers are of particular interest, because, on the one hand, these are particularly relevant to general Twitter users—as opposed to, for example, physics research—and practicing physicians belong to early adopters of social media in their work practice (Berger, 2009). In a survey asking researchers about social media use in research, the uptake by health scientists was, however, slightly below average (Rowlands, Nicholas, Russell, Canty, & Watkinson, 2011).

The aim of this paper is thus to investigate whether there is a connection between different medical study types and the frequency of tweets per article. We hypothesize that some study types are more popular on Twitter due to their attractiveness for a broader audience such as applied medical research relevant to patients as well as meta-analyses summarizing research and condensing results. We will approach this hypothesis by first investigating the potential differences in tweet frequency for a range of medical study types. We argue that logically there should be a connection between the clinical evidence hierarchy (further explained below) and the types of studies patients might consider interesting to discuss or spread on social media, as the highest evidence levels are those which are most likely to affect clinical practice. We therefore expect differences in tweet frequency to be related to evidence levels.

**Materials and Methods**
Comparing the impact of medical research study types on Twitter requires two pieces of information per research article: a classification of the study type as well as the number of tweets received by each particular paper. Currently no database contains both pieces of information, so that it was necessary to combine data from different sources. For this purpose, the medical study type classifications from the Embase bibliographical database was used, enriched with metadata from PubMed and Web of Science and then matched to Twitter data from Altmetric.com. The datasets and the matching approach are described in further detail below. Following these descriptions is an account of the specific measurements and statistical tools employed as well as the limitations of this study.

*Data collection and matching*

Due to Twitter's 140 character limitation, mentions of a scientific paper in tweets are restricted to links to the publisher's homepage or unique document identifiers such as the Digitial Object Identifier (DOI) or PubMed ID (PMID). As Twitter only provides access to the most recent tweets[2], it is necessary to constantly query various article identifiers to obtain a database of tweets to scientific papers. Altmetric LLP has been collecting tweets based on multiple document identifiers including the DOI, PMID and the publisher's URL since July 2011 and thus provides a valuable data source for the purposes of our study. To assure reliable and complete Twitter data, we focus our study on papers published 2012. In order to link all tweets to the bibliographic data and study type classification from Embase, the DOI and the PMID are needed.

The study type classifications (see below) for the analysis were retrieved from the Embase bibliographical database. Embase is a major database containing more bibliographical records than PubMed Medline; for example, 24%[3] more for documents published in 2012. It is unclear whether the study type classifications of either database outperforms the other, however, as the indexing of Embase is more exhaustive, we have chosen to use this database for our study. In order to identify relevant papers from Embase (and to be able to perform a citation analysis in the future), *Clinical Medicine* journals were selected from the Web of Science (WoS) based on the National Science Foundation (NSF) journal classification system. The Web of Science also provides bibliographic data and DOIs for the relevant papers which were used to match Embase study types and tweets from Altmetric.

Embase was queried for the relevant journals using the journal name and various abbreviations as well as the ISSN. Limiting the results to papers published in 2012, the metadata of 593,974 records was retrieved from Embase. In order to obtain the PMID needed to match tweets, PubMed was queried in the same way resulting in 497,619 records. Embase, PubMed and Web of Science were matched using the DOI, PubMed as well as string matches of bibliographic information resulting in 238,560 documents in the final dataset, 94.9% of which with a PMID and 91.1% with a DOI.

The bibliographic metadata was matched to the Altmetric database using the DOI and PMID resulting in 80,116 records with at least one social media event as captured by Altmetric and 74,060 with at least one tweet at the time of data collection in August 2014. This amounts to 31% of the 238,560 being mentioned on Twitter at least once, which corresponds almost exactly to the Twitter coverage of biomedical & health sciences papers found by Haustein, Costas and Larivière (2015). To ensure comparability between tweets published in January and December 2012, we fixed the tweeting window to 18 months (546 days) for each of the tweeted documents, including tweets until 30 June 2013 for papers published on 1 January 2012 and until 30 June 2014 for papers published on 31 December 2012. The day of publication is based on the publication date provided by Altmetric. As this date is not available for all records and is sometimes incorrect, the dataset was further reduced to 52,911 documents, which had an Altmetric publication date in 2012 and not received a tweet before the publication date. Although these steps lead to an underestimate of the percentage of tweeted papers, they help to reduce biases induced by publication age when comparing the visibility of different medical study types on Twitter.

---

[2] Twitter's REST API is limited to tweets from the previous week, while the Streaming API provides realtime data only.
[3] For the publication year 2012, Embase contains 1,334,356 records (search: "2012".yr) and PubMed Medline contains 1,072,384 (search: 2012[pdat]).

*Medical study type classification*

Embase indexes all articles using a controlled vocabulary (the Emtree thesaurus), which contains hierarchically ordered keywords in a classical thesaurus structure. Among these keywords are study type classifications, of which some are directly identifiable as such (e.g. randomised controlled trials), while others require some translation (e.g. "sensitivity and specificity" which is used for diagnostic accuracy studies). The Emtree thesaurus is designed for indexing and retrieval, and there is thus not a given connection between the hierarchical ordering of study type keywords and different levels of research methodology. This is particularly important, as one of the predominant approaches to Western medical research and practice is the so-called evidence based medicine (EBM). One of the cornerstones of EBM is the distinction between study types and their hierarchical ordering based on how much 'evidence' a study is assumed to contribute to the understanding of a given problem (Greenhalgh, 2010). Different hierarchies exist, e.g. the Oxford Centre for Evidence Based Medicine's "Levels of Evidence" (OCEBM Levels of Evidence Working Group, 2011).

**Table 1. Medical study type classification system based on Röhrig et al (2009) and OECBM. Classifications with raised numerals have narrower terms which are not shown here.**

| | Medical research | | | | | | | |
|---|---|---|---|---|---|---|---|---|
| | Primary research | | | | | | Secondary research | |
| research_type | A. Basic research | | B. Clinical research | | C. Epidemiological research | | D. Synthesising research | |
| class | A1. Theoretical | A2. Applied | B1. Experimental | B2. Observational | C1. Experimental | C2. Observational | D1. Meta-analysis | D2. Review |
| | Method development | Animal study; cell study; genetic engineering/sequencing; biochemistry; material development; genetic studies | Clinical study; phase I-IV | Therapy; prognostic; diagnostic; observational study with drugs; secondary data analysis; case series; case report | Intervention study; field study; group study | Cohort (prospective/historical); case control; cross-sectional; ecological; monitoring, surveillance; Description with registry data | | Systematic; narrative |
| study_type embase_keyword | A1.1 Theoretical study *Theoretical study*  A1.2 Method development *-* | A2.1 Ex vivo study *Ex vivo study*  A2.2 In vivo studies *Animal experiment*  A2.3 In vitro study *Animal tissue, cells or cell components[i] Cell, tissue or organ culture[ii] Human tissue, cells or cell components[iii]*  A2.4 Genetic engineering *Genetic engineering and gene technology Genetic engineering Gene sequence*  A2.5 Biochemistry *Biochemistry Neurochemistry Phytochemistry* | B1.1 Clinical trial *Clinical trial Clinical trial (topic) Controlled clinical trial Multicenter study Phase 1 clinical trial Phase 2 clinical trial Phase 3 clinical trial Phase 4 clinical trial Randomized controlled trial* | B2.1 Case study *Case report Case study*  B2.2 Prognostic study *Prognosis*  B2.3 Diagnostic study *Diagnosis Diagnostic test Sensitivity and specificity*  B2.4 Therapy *-*  B2.5 Observational study with drugs *Observational study AND (major clinical study OR controlled study OR clinical article)* | C1.1 Intervention study *Intervention study*  C1.2 Field study *Field study*  C1.3 Group study *-* | C2.1 Case control study *Case control study[iv]*  C2.2 Cohort study *Cohort study Longitudinal study Retrospective study Prospective study*  C2.3 Cross sectional study *Cross-sectional study*  C2.4 Ecological study *-*  C2.5 Monitoring *Patient monitoring*  C2.6 Surveillance *-*  C2.7 Registry study *-* | D1.1 Meta-analysis *Meta-analysis* | D2.1 Review *Review Systematic review* |

We have chosen to use a particular hierarchy which allows a classification of study types on their level of research (Röhrig et al., 2009). We have added to the classification of Röhrig et al. (2009) by adding classification codes and the corresponding keywords in Emtree. The resulting system has been validated by two field-experts, and is displayed in Table 1. As can be seen, the classification system allows direct translation between specific Emtree keywords (we have added the broadest terms as well as their relevant narrower terms) and our classification codes on the third level (study_type). The system allows grouping of study types into classes and research types (levels 2 and 1), thus allowing us to analyse the connection between tweets and the specific study types as well as the broader categories.

Of the entire population of 238,560 records, 162,830 records can be classified using our study type classification system. Of these, 36,595 (22.5%) receive at least one tweet within the fixed 18 months tweet window. Of the remaining 75,730 records without a classification, 16,316 (21.5%) receive at least one tweet. These data delimitations will be used to control for systematic errors in our main dataset (records with classifications). Among those that were classified, 55% had only one classification, 26% had two, 12% had three and the remaining 7% had four or more classifications. References with $n$ classifications are treated as $n$ observations, thus resulting in more than 162,830 observations on either classification level. Some classes in our classification system were not observed at all in the dataset. These classes are omitted in the results section.

*Statistical methods and indicators*

For each study type classification level we report several statistics for all documents (referred to by $*_A$, e.g. $N_A$) as well as the subset that has received at least one tweet ($*_T$). The included statistics are number of articles per classification ($N$), mean tweets per article ($\mu$), the standard deviation from the mean ($\sigma$), percentage of articles with at least one tweet ($N_T/N_A$), and the mean normalised tweets ($\hat{\mu}$) defined as the ratio between $\mu$ for a specific classification and $\mu$ for the entire population.

As the distributions of tweets for any classification are extremely skewed (see results) similar to citations, the adequacy of the mean as an indicator of average activity is debatable (Calver & Bradley, 2009). However, while the median might be a methodologically more sound choice, the distributions are so extremely skewed that for study type level classification, medians are all 0 when all papers are included and either 1 or 2 if only tweeted papers are included. The corresponding means range from 0.35 to 1.74 and 2.02 to 5.01, providing considerably more information, especially as the scales for the mean are continuous. We therefore use the mean for comparisons, with due care and inclusion of standard deviations and percentage of tweeted articles to provide further information on differences in means. As we have large sample sizes, we expect any major differences in means to be real and not due to chance. However, to test this assumption, all classifications are tested pairwise and against the background population using the independent sample, unpaired Mann-Whitney test.

*Limitations*

The most obvious error source in this study is the proportion of papers included in the final analysis, compared to the overall population of papers published in 2012. Our background population of 162,830 classified papers only represents 27.4% of the 593,974 records downloaded from Embase. However, it still represents 68.3% of the 238,560 matchable records. This is a fairly high number of papers that could be classified, and if it is possible to improve the matching algorithms, it should also be possible to increase the total number of classified papers comparably. The only systematic error in this regard is the omission of particular documents based on lacking or erroneous DOI's. However, as missing DOI's are also an issue in collecting tweets, this error is not likely to affect the tweet counts with the limitations to tweet-collection that currently exist.

To test if there is a systematic error in the number of tweets per paper, with regard to whether a paper has been classified with a study type or not, we compare the percentage of papers with tweets for classified papers with unclassified papers. For the 162,830 papers with a classification, 36,595 (22.5%) received at least one tweet, while the 75,730 unclassified papers received tweets on 16,316 (21.5%) papers. These values also corroborate findings by Haustein, Costas & Larivière (2015). For the classified papers, mean tweets were 0.67, while the mean was 0.71 for the unclassified papers. These differences are not random (p = 2.7e-14, using independent two-sample t-test), however, the effect size is also extremely small (Cohen's $d$ = 0.018). We should therefore not consider the lack of study types as confounders for the number of tweets.

While the classification system we have used here was validated by two domain experts, it is only one possible system. Other classifications could have been created, in particular with regard to the translation from Emtree keywords to our classification system. The choices made in this regard will affect the results as presented here. However, when we compare the pairwise scores within a research class, we find high consistency between what could be considered "similar" research types. The only study type which varies greatly from the other study types in their class is the non-systematic review. This is meaningful, as non-systematic reviews are regarded by medical researchers as much less evidential as their systematic counterparts.

**Results**

We analysed the classified papers on the three levels present in our classification system: research type, research class and study type. In Tables 2 to 4 we report summary statistics for the three levels, for all papers as well as limited to tweeted papers to determine differences between the share of tweeted papers as well as intensity of (re)use. Results are visualized in Figure 1. In Figures 2 to 4 we provide the results of the pairwise comparison to determine the statistical significance of differences between study types including binary and continuous statistical significance as well as Cohen's d to estimate effect size.

*Summary statistics*

As can be seen from Tables 2 to 4, there are large differences in the mean tweets per classification, regardless of classification level, although the largest differences are observable in the study types. The differences are clear from the means ($\mu_A$ and $\mu_T$), but even more obvious when regarding the relative means ($\hat{\mu}_A$ and $\hat{\mu}_T$). This is also where we find the largest standard deviations, likely due to the smaller *N* per classification. Meta-analyses and systematic reviews receive considerably more tweets than other study types, which makes the synthesizing research type stand out as well. Overall, a generally increasing interest of the Twitter community can be observed from basic (A) over clinical (B) and epidemiological (C) to synthesizing research (D) papers. Larger variations per research type can be observed for clinical research, where clinical trials are much more tweeted than other study types. In fact, case studies (B2.1) have the lowest mean number of tweets per paper ($\mu_A$), which also reflects in the low mean of observational clinical research (B2) on the research class level. Epidemiological research also performs above average of the entire sample, while basic research (A) consequently performs below, although with somewhat higher scores for genetic engineering (A2.4) than the papers classified as ex vivo (A2.1), in vivo (A2.2) and in vitro (A2.3) studies.

**Table 2. Summary statistics for research type.**

| Research type | $N_A$ | $\mu_A$ | $\sigma_A$ | $N_T$ | $N_T/N_A$ | $\mu_T$ | $\sigma_T$ | $\hat{\mu}_A$ | $\hat{\mu}_T$ |
|---|---|---|---|---|---|---|---|---|---|
| A. Basic research | 130,171 | 0.434 | 1.491 | 25,992 | 0.200 | 2.172 | 2.712 | 0.642 | 0.743 |
| B. Clinical research | 70,262 | 0.766 | 2.699 | 16,623 | 0.237 | 3.238 | 4.773 | 1.133 | 1.108 |
| C. Epidemiological research | 43,733 | 0.963 | 3.201 | 12,132 | 0.277 | 3.472 | 5.313 | 1.425 | 1.188 |
| D. Synthesising research | 38,558 | 1.005 | 3.223 | 10,641 | 0.276 | 3.640 | 5.295 | 1.486 | 1.245 |

**Table 3. Summary statistics for research class.**

| Research class | $N_A$ | $\mu_A$ | $\sigma_A$ | $N_T$ | $N_T/N_A$ | $\mu_T$ | $\sigma_T$ | $\hat{\mu}_A$ | $\hat{\mu}_T$ |
|---|---|---|---|---|---|---|---|---|---|
| A2. Applied basic research | 130,171 | 0.434 | 1.491 | 25,992 | 0.200 | 2.172 | 2.712 | 0.642 | 0.743 |
| B1. Experimental clinical research | 28,343 | 1.219 | 3.495 | 8,949 | 0.316 | 3.860 | 5.337 | 1.803 | 1.321 |
| B2. Observational clinical research | 41,919 | 0.460 | 1.928 | 7,674 | 0.183 | 2.511 | 3.894 | 0.680 | 0.859 |
| C2. Observational epidemiological research | 43,733 | 0.963 | 3.201 | 12,132 | 0.277 | 3.472 | 5.313 | 1.425 | 1.188 |
| D1. Meta-analyses | 1,883 | 1.742 | 4.488 | 655 | 0.348 | 5.009 | 6.448 | 2.577 | 1.714 |
| D2. Reviews | 36,675 | 0.967 | 3.139 | 9,986 | 0.272 | 3.550 | 5.199 | 1.430 | 1.215 |

**Table 4. Summary statistics for study type.**

| Study type | $N_A$ | $\mu_A$ | $\sigma_A$ | $N_T$ | $N_T/N_A$ | $\mu_T$ | $\sigma_T$ | $\hat{\mu}_A$ | $\hat{\mu}_T$ |
|---|---|---|---|---|---|---|---|---|---|
| A2.1. Ex vivo study | 1,061 | 0.425 | 1.285 | 223 | 0.210 | 2.022 | 2.155 | 0.629 | 0.692 |
| A2.2. In vivo study | 52,127 | 0.437 | 1.435 | 10,676 | 0.205 | 2.135 | 2.536 | 0.647 | 0.731 |
| A2.3. In vitro study | 75,287 | 0.427 | 1.519 | 14,699 | 0.195 | 2.190 | 2.821 | 0.632 | 0.749 |
| A2.4. Genetic engineering | 1,696 | 0.606 | 1.951 | 394 | 0.232 | 2.607 | 3.345 | 0.896 | 0.892 |
| B1.1. Clinical trial | 28,343 | 1.219 | 3.495 | 8,949 | 0.316 | 3.860 | 5.337 | 1.803 | 1.321 |
| B2.1. Case study | 21,788 | 0.348 | 1.847 | 3,204 | 0.147 | 2.367 | 4.292 | 0.515 | 0.810 |
| B2.2. Prognostic study | 6,618 | 0.525 | 1.842 | 1,407 | 0.213 | 2.469 | 3.341 | 0.776 | 0.845 |
| B2.3. Diagnostic study | 13,513 | 0.608 | 2.081 | 3,063 | 0.227 | 2.682 | 3.680 | 0.899 | 0.917 |
| C2.1. Case control study | 2,428 | 0.975 | 3.547 | 664 | 0.273 | 3.566 | 6.065 | 1.443 | 1.220 |
| C2.2. Cohort study | 34,822 | 0.943 | 3.163 | 9,585 | 0.275 | 3.424 | 5.276 | 1.394 | 1.171 |
| C2.3. Cross sectional study | 4,891 | 1.106 | 3.300 | 1,440 | 0.294 | 3.756 | 5.201 | 1.636 | 1.285 |
| C2.5. Monitoring | 1,592 | 0.956 | 3.163 | 443 | 0.278 | 3.436 | 5.242 | 1.414 | 1.175 |
| D1.1. Meta-analysis | 1,883 | 1.742 | 4.488 | 655 | 0.348 | 5.009 | 6.448 | 2.577 | 1.714 |
| D2.1. Review | 32,962 | 0.885 | 2.909 | 8,694 | 0.264 | 3.354 | 4.878 | 1.309 | 1.147 |
| D2.2. Systematic review | 3,713 | 1.695 | 4.653 | 1,292 | 0.348 | 4.871 | 6.839 | 2.507 | 1.666 |

The distributions of tweets per classification are shown in Figure 1, illustrating the highly skewed nature of these distributions, but also the large differences between some categories. The results shown in these boxplots are directly comparable to the summary statistics, and the same classifications stand out as being particularly often tweeted.

From previous research we know that meta-analyses, systematic reviews and clinical trials are also the most highly cited study types (Andersen & Schneider, 2011). However, whether there is a connection between the citedness and tweetedness of medical study types is not obvious from the present data, and will require further research.

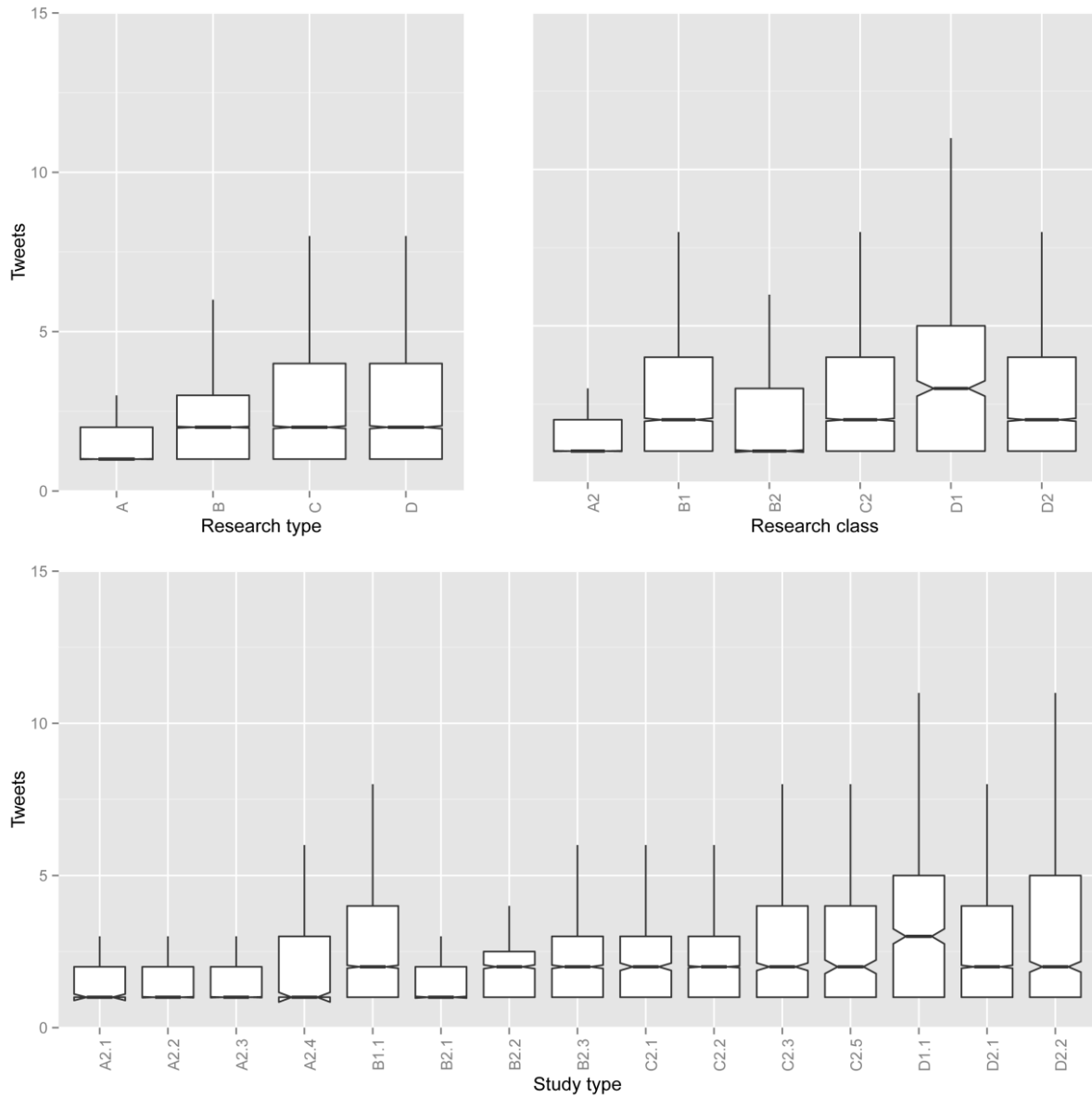

**Figure 1. Notched boxplots showing tweet distributions for A) Research type, B) Research class and C) Study type.**

*Pairwise comparison*

In order to analyse the magnitude of differences in classifications further, pairwise comparisons were made on each level. The independent two-sample Mann-Whitney test was used to test whether differences in sample means were due to random effects, and Cohen's $d$ was used to estimate the effect size of varying means. There is of course a connection between the $p$-values of the Mann-Whitney tests and Cohen's $d$, to the extent that non-significant differences will also have very small effect sizes, as our sample sizes are quite large. In Figures 2 to 4 these pairwise comparisons are plotted as heatmaps, in which the diagonal and lower half have been omitted. The statistical significance of differences in mean are plotted as both binary maps ($p$ below or above 0.05) and as continuous values. On the research type level, basic research stands out the most from the other types, with a lower mean of tweets per paper. For research classes, meta-analyses stand out with very large effect sizes, but overall the effect sizes are somewhat larger on this level than the broader research types. On the study type level, meta-analyses and systematic reviews stand out, but also clinical trials and epidemiological study types have fairly large effect sizes, compared to other study types.

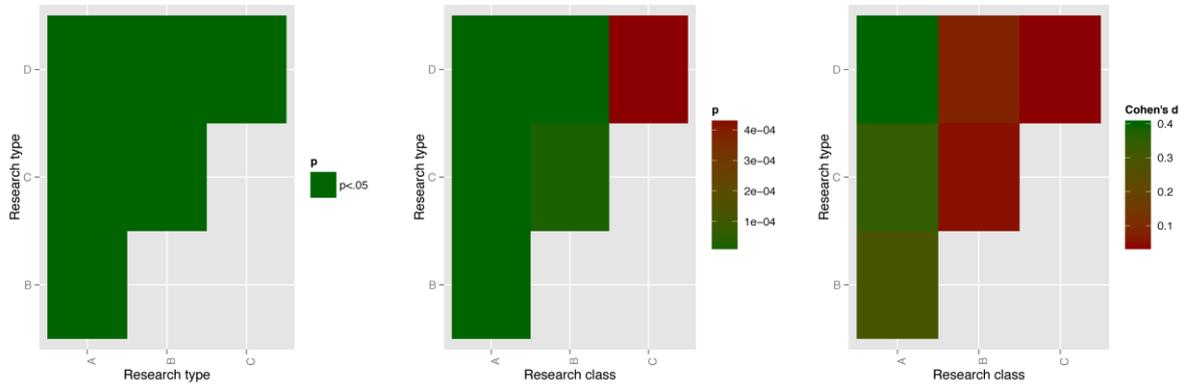

**Figure 2. Heatmaps of pairwise comparisons showing A) binary statistical significance, B) continuous statistical significance and C) Cohen's *d* as effect size estimate. All figures are grouped on the research type level.**

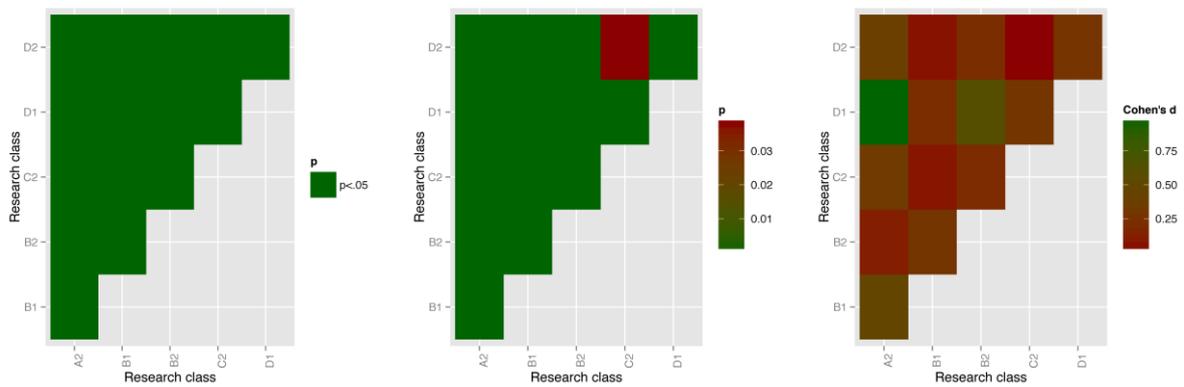

**Figure 3. Heatmaps of pairwise comparisons grouped on the research class level. See figure 2 for legend.**

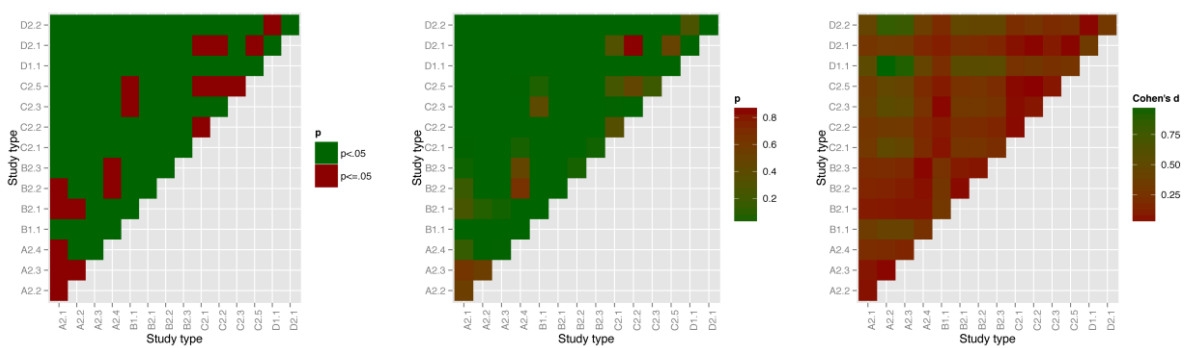

**Figure 4. Heatmaps of pairwise comparisons grouped on the study type level. See figure 2 for legend.**

**Discussion and Outlook**

We have analysed the frequency of tweets for medical research papers, distinguished by their specific study type. Our hypothesis was that some study types would be more frequently tweeted, because they were interesting to a wider audience (e.g., patients and other laymen) than other types. It has not been possible to identify literature on which types of research are actually more useful to laymen, or even which types are most often used. We therefore assume that research which is close to clinical practise and may contribute to changes in treatments would be more interesting to patients, as they might see a specific benefit to themselves. Based on findings by Haustein, Costas and Larivière (2015) that briefer and condensed document types received more tweets than research articles, we also assumed that synthesising research papers would be more popular on Twitter than basic research.

On the broadest classification level, the results fit well with this assumption, as basic research stands out as the least frequently tweeted research type on average. Basic medical research is also furthest removed from the actual treatment of diseases—so much that some physicians consider it irrelevant to their clinical practise (Andersen, 2013)—which makes them less interesting for the general public of medical laymen and patients active on Twitter. When fine-tuning the analysis to study types, meta-analyses and systematic reviews stand out particularly, followed by clinical trials and epidemiologic study types. This corresponds with typical evidence hierarchies and reflects similar patterns found for citations (Andersen & Schneider, 2011; Kjaergard & Gluud, 2002; Patsopoulos et al., 2005). While this might indicate a relationship between tweets and citations, other studies on a broader level have found this is not the case (Costas et al., 2014; Haustein et al., submitted; Haustein, Larivière, et al., 2014; Zahedi et al., 2014). Other explanations may be that physicians are more likely to tweet about high-evidence studies or that these are also the same types of studies which are most interesting to patients. The latter appears obvious, as high-evidence studies are also more likely to be included in clinical practice guidelines and thus have a greater potential for changing practice. Moreover, results indicating the uptake of social media to be lower among health researchers (Rowlands et al., 2011), while the frequency of tweets per paper in this area is high (Haustein, Peters, et al., 2014), provide some evidence, that the large effect size found for these study types cannot be explained purely by large Twitter-activity from medical researchers. Patients, patient groups and laymen interested in research or other factors may thus play an important role in this observation.

While factors such as entertaining topics may play a role (Neylon, 2014) when looking at the the top per mille most frequently tweeted papers, it is unlikely that all 1,883 meta-analyses, 3,713 systematic reviews and 28,343 clinical trials should have a higher tweet count than other study types due to entertainment value, especially as these are also the most highly regarded study types by the researchers as measured through citations. The mean may of course be affected by single high-scoring studies, however, as can be seen from Figure 1, it is the entire distribution rather than merely the mean which is increased for these study types. In fact, the maximum tweets per study type is 46 for meta-analyses and 59 for systematic reviews, while it is 65 for two of the basic research study types and 62 for clinical trials. The lowest maximum tweet frequency of a study type is 25 (an in vivo study) and the highest is 67 (a cohort study). It can thus be concluded that medical study types are one of the factors determining popularity of scientific papers on Twitter but they are certainly not the only ones. Apart from factors explored by previous studies and known also from the citation context—such as discipline, publication age, number of authors etc.—Twitter-specific effects should also be investigated. This includes the effect of the number of followers and affordance use as well as the extent to which scientific papers receive tweets due to author and journal self-promotion as well as automated Twitter accounts (Haustein, Bowman, et al., 2015).

**Acknowledgements**

The authors would like to thank Euan Adie and Altmetric.com for access to their Twitter data. SH acknowledges funding from the Alfred P. Sloan Foundation, grant no. 2014-3-25.